\documentclass[final,1p,times]{elsarticle}

\usepackage{amssymb}
\usepackage{amsmath}
\usepackage{lineno}

\newcommand{\be}{\begin{eqnarray}}
 \newcommand{\eqn}[1]{Eq.\,(\ref{#1})}

\newcommand{\ee}{\end{eqnarray}}

\newcommand{\ave}[1]{\left\langle #1 \right\rangle}

\begin{document}

\begin{frontmatter}



\title{Hydrodynamic Initial Conditions in Small Systems from Proton Phase-Space Entropy} 


\author[label1]{Gabriel Rabelo-Soares}
 \ead{rabelosoares21@gmail.com}

\author[label2]{Gojko Vujanovic}
\ead{gojko.vujanovic@uregina.ca}

\author[label1]{Giorgio Torrieri}
\ead{torrieri@ifi.unicamp.br}

\affiliation[label1]{organization={Universidade Estadual de Campinas - Instituto de Física Gleb Wataghin},
            addressline={Rua Sérgio Buarque de Holanda, 777}, 
            city={Campinas},
            postcode={13083-859}, 
            state={Sao Paulo},
            country={Brazil}}

\affiliation[label2]{organization={Department of Physics, University of Regina},
            addressline={3737 Wascana Parkway}, 
            city={Regina},
            postcode={S4S 0A2}, 
            state={Saskatchewan},
            country={Canada}}

\begin{abstract}
The experimental observation of collective behaviour in proton-proton and proton-nucleus collisions poses a fundamental theoretical question regarding the proper characterization of the initial state underlying hydrodynamic evolution. While relativistic hydrodynamics requires an initial condition (IC) characterized by an entropy current, corresponding to a maximally mixed state, the microscopic description of the proton is based on inherently quantum objects, that are projections of pure states. We show that the appropriate matching between proton wave function and classical hydrodynamics emerges from the coarse-graining of its phase-space distribution quantified by the Wehrl-\textit{like} entropy. This entropy provides a semi-classical, positive-definite measure of the density of accessible microstates at a given resolution scale, and therefore constitutes the appropriate quantity to characterize entropy deposition in small collision systems.
\end{abstract}




\end{frontmatter}





\section{Introduction}
\label{sec1}

For over two decades, the ultra-relativistic heavy-ion program at RHIC and the LHC has investigated the properties of the quark-gluon plasma (QGP), a phase of quantum chromodynamics in which quarks and gluons are deconfined. Experimental and theoretical studies indicate that this state of matter behaves as an almost perfect fluid \cite{heinz_collective_2013}, motivating the successful application of relativistic hydrodynamics to nucleus-nucleus (\textit{AA}) collisions. 

More recently, experimental observations have also revealed signatures of collectivity in small collision systems \cite{nagle,atlas_collaboration_measurement_2018}. This observation has led to the extension of hydrodynamic modelling to proton-proton and proton-nucleus collisions, where one of the central open questions is how to properly characterize the initial conditions of the hydrodynamic evolution.

In \cite{Rabelo-Soares:2025dfu} we address the problem of characterizing the initial conditions of hydrodynamic evolution in small collision systems. Since the transverse size of the system is of the order of the proton size,
the initial state is dominated by quantum fluctuations. Matching this quantum description to a deterministic hydrodynamic framework therefore requires a coarse-grained characterization of the proton quantum state. In this
work we explore a scenario in which the hydrodynamic initial condition is determined by the proton phase-space entropy, described through a Wehrl-\textit{like} entropy obtained from the coarse-grained proton quantum state. 


\section{Phase-Space Description of the Proton}

The proton wave function can be described in light-front quantization as a superposition of partonic Fock states. Since the hadron state is pure, the corresponding density matrix has vanishing Shannon entropy. The Wigner distribution characterizes the phase-space structure of the proton, which depends on the longitudinal momentum
fraction $x$, the intrinsic transverse momentum $\mathbf{k}_\perp$, and the transverse position $\mathbf{b}_\perp$ of the partons.

Because of the uncertainty principle, the Wigner distribution cannot be interpreted as a probability density in phase-space. A positive semi-classical
distribution can be obtained through Gaussian smearing of the Wigner function, the Husimi distribution, which is defined as 
\begin{equation}
\label{wherldef}
H_{\ell}(x, \mathbf{r}, \mathbf{p}) =
\frac{1}{\pi^{2}}
\int d^{2} \mathbf{r}^{\prime} d^{2} \mathbf{p}^{\prime}
\exp\left[-\frac{(\mathbf{r}- \mathbf{r}^{\prime})^{2}}{\ell^{2}}\right]
\exp\left[-(\mathbf{p} - \mathbf{p}^{\prime})^{2} \ell^{2}\right]
W(x,\mathbf{r}^{\prime}, \mathbf{p}^{\prime}),
\end{equation}
where $\ell$ is a parameter with units of length associated with the
resolution scale of the system. As will be discussed later, $\ell$ will be related to the thermalization scale $\Lambda$ via $\ell \sim \Lambda^{-1}$. In this case
$\langle \ln H_\ell \rangle$ is non-zero even for a pure state, since $H_\ell$ represents a semiclassical distribution obtained from a coarse-grained description of the phase-space.

\section{Hydrodynamic Initial State\label{initialhydro}}

Considering that hydrodynamics is indeed applicable in small collision systems, the relevant question becomes how to relate the microscopic quantum structure of the proton to the macroscopic quantities that determine the hydrodynamic initial state. In a scenario where the initial state is already thermalized, i.e., microstates are localized and equally likely, entropy current provides a more appropriate macroscopic measure of the available microstates in the phase-space. 

A relevant initial state should behave as a maximally decohered state. Experimentally, this has been verified \cite{kharzeev3} for point-like deep inelastic scattering events and $pp$ collisions, where the entropy inferred from the parton wave function of the nucleon matches the measured value $\ave{\ln P(N)}$ with $P(N)$ the distribution of multiplicities.

To construct such a quantity, it is convenient to express the proton structure in terms of its partonic degrees of freedom. A natural framework for this purpose is provided by the light-cone wave function (LCWF) formalism, in which the proton state is expanded in a superposition of partonic Fock states.

Within this description, the gluon structure function can be written as the probability of finding a gluon carrying longitudinal momentum fraction $x$ at a resolution scale $Q^2$. By Fourier transforming the transverse momenta to transverse coordinates and using Parseval's theorem, the gluon distribution can be expressed as

\begin{equation}
\label{gdef}
G(x,Q^2)= \sum_{n,\lambda_i} 
\int \prod dx_i d^{2} \mathbf{r}_{\perp i}
\left| \tilde{\Psi}^Q\left(\mathbf{r}_{\perp i},x_i,\lambda_i \right)\right|^{2}
\sum_i\delta\left(x-x_i\right),
\end{equation}
where $\tilde{\Psi}^Q$ is the Fourier transform of $\Psi^Q\left(\mathbf{k}_{\perp i},x_i,\lambda_i\right)$ in the LCWF. Counting the partons in a transverse cell $d^{2} \mathbf{b}_\perp $ one obtains
\begin{equation}
\label{rhodef}
\bar{G}(x,\mathbf{b}_\perp,Q^{2})= \sum_{n,\lambda_i} \int \prod dx_i d^{2} \mathbf{r}_{\perp i}\left| \tilde{\Psi}^Q\left(\mathbf{r}_{\perp i},x_i,\lambda_i \right)\right|^{2}\sum_i\delta\left(x-x_i\right)\delta^{(2)}\left( \mathbf{r}_\perp - \mathbf{b}_{\perp i}\right).
\end{equation}
Like \eqn{gdef}, \eqn{rhodef} is independent of the phase of the gluon fields, and hence according to \cite{kharzeev} it qualifies as a maximally decohered quantity which can serve as a basis for an initial entropy density. The $Q^{2}$ dependence of \eqn{rhodef}, just like in \eqn{gdef}, comes from renormalization group running of LCWF, and is therefore $\sim \ln Q^{2}$.

\section{From the Proton Phase-Space Distribution to Initial Conditions} \label{wignerentropycontent}
From Wigner function, one may quantify quantum coherence through entropy-like measures. For a positive Husimi distribution the Wehrl entropy
\cite{hattawherl} provides a natural probe of the
phase-space structure of the proton.

In quantum field theory the number of degrees of freedom grows
with resolution. As a consequence the Wehrl entropy diverges in
the ultraviolet limit, $\langle \ln \mathcal{W} \rangle \sim \Lambda^3$,
reflecting the proliferation of modes. A physical entropy can
therefore only be defined after introducing a resolution scale
$\Lambda$, which acts as a coarse-graining scale.

For an extended system where local equilibrium and hydrodynamics are expected to be relevant, $\Lambda$ is determined by the energy budget available in the local thermalized cell. The spatial size of this cell ($L$) constrains the accessible momentum scale through the uncertainty principle ($\Lambda \sim L^{-1}$), 
\begin{equation}
\label{wherlqft}
\mathcal{W}_{\Lambda}(\phi,\Pi)= \int  \mathcal{D}\phi' \mathcal{D}\Pi' \exp\left[-\Lambda^{2} (\phi-\phi')^{2}\right] \exp \left[-\frac{(\Pi-\Pi')^{2}}{\Lambda^{2}}\right] \mathcal{W}(\phi',\Pi'),
\end{equation}
the localization scale $\Lambda$ hence enters into the Gaussian smearing, leading to the entropy density $\sim \Lambda^{3}$ as required; $\ave{\ln \mathcal{W}_\Lambda}$ is an estimate for the entropy of a field whose degrees of freedom are localized in cells of size $\Lambda^{-1}$.

Connecting \eqn{wherlqft} to the light-cone distributions described in the previous section can be done through the formalism of \cite{Belitsky:2003nz} together with the definition of Wehrl entropy. In this picture, probing the proton at a scale $Q^{2}$ effectively breaks the quantum coherence of the partonic state at that scale, increasing the number of accessible microstates. This leads to the relation
\begin{equation}
\ave{\ln \mathcal{W}_\Lambda}\left(x,\mathbf{b}_\perp\right) =
\mathcal{N}\int_{0}^{\Lambda} d^{2}Q\,\bar{G}(x,\mathbf{b}_{\perp},Q^{2}).
\label{qq}
\end{equation}
The physical interpretation of this equation is that the localization of degrees of freedom described by the probability distributions in \eqn{rhodef} is a good proxy for the Husimi smearing of the Wigner functional of the coherent hadronic state.  Probing parton degrees of freedom at $Q^{2}$ breaks the coherence of the Wigner functional at that scale, producing  an amount of entropy approximated by the Husimi distribution. This entropy reflects the information loss associated with the coarse-grained Husimi distribution.

Once an entropy density is defined following the previous description, it can be used as the initial entropy profile entering the hydrodynamic evolution. Given a model for the Wigner function of the nucleon, the total multiplicity of a \textit{pp} or a \textit{pA} collision is sufficient to set the scale $\Lambda$ of the entropy. Since entropy is by its nature extensive, the target and projective entropies will add up in such way to form the entropy deposition of the collision. Therefore one can calculate the eccentricities of the entropy distribution, and hence it be used as an estimate of the $v_n$ in $pp$ and $pA$ collisions as calculated from an initial Wigner function in order to compare with the available data. Quantitative analysis is left for a future work.

\section{Discussion and Conclusions \label{discussion}}
In this work we explored a possible connection between the quantum phase-space structure of the proton and the hydrodynamic initial conditions from the Wehrl-\textit{like} entropy. Quantum mechanics and quantum field theory do not admit such a description due to the intrinsic contextuality of quantum expectation values. To obtain quantitative predictions for the hydrodynamic response in \textit{pp} and \textit{pA} collisions, one must therefore establish a consistent matching between the two pictures.

In conclusion, we explored the relationship between the initial state of the fluid and the objects usually used to characterize the structure of the nucleon. The initial state and the partonic distributions are not directly linked in the sense that the initial state of hydrodynamics in small systems will not be a straight-forward function of such objects. The full implications of this issue require further analysis and are not addressed herein. Nevertheless, a comparison of the flow harmonics $v_n$ in \textit{pA} and \textit{pp} collisions with polarized protons might provide a quantitative probe of this three-dimensional structure.

\section*{ACKNOWLEDGMENTS}
G.R-S. is supported by the CAPES doctoral fellowship 88887.005836/2024-00. G.T. thanks Bolsa de produtividade CNPQ 305731/2023-8 and FAPESP 2023/06278-2, as well as FAPESP temático 2023/13749-1 for support. G.V. is grateful for the support provided by the Canada Research Chair under grant number CRC-2022-00146 and the Natural Sciences and Engineering Research Council (NSERC) of Canada under grant number SAPIN-2023-00029.







\bibliographystyle{elsarticle-num} 
\bibliography{ref}






\end{document}